Exploring the Psychological Basis for Transitions in the Archaeological Record

Liane Gabora
Department of Psychology, University of British Columbia
Fipke Centre for Innovative Research, 3247 University Way,
Kelowna, BC Canada, V1V 1V7
liane.gabora@ubc.ca
https://people.ok.ubc.ca/lgabora/

Cameron M. Smith
Department of Anthropology, Portland State University Portland, OR, 97207
b5cs@pdx.edu
https://www.pdx.edu/anthropology/cameron-smith

Exploring the Psychological Basis for Transitions in the Archaeological Record

## Introduction

No other species remotely approaches the human capacity for the cultural evolution of novelty that is accumulative, adaptive, and open-ended (i.e., with no a priori limit on the size or scope of possibilities). By *culture* we mean extrasomatic adaptations—including behavior and technology—that are socially rather than sexually transmitted. This chapter synthesizes research from anthropology, psychology, archaeology, and agent-based modeling into a speculative yet coherent account of two fundamental cognitive transitions underlying human cultural evolution that is consistent with contemporary psychology. While the chapter overlaps with a more technical paper on this topic (Gabora & Smith 2018), it incorporates new research and elaborates a genetic component to our overall argument. The ideas in this chapter grew out of a non-Darwinian framework for cultural evolution, referred to as the *Self-other Reorganization (SOR)* theory of cultural evolution (Gabora, 2013, in press; Smith, 2013), which was inspired by research on the origin and earliest stage in the evolution of life (Cornish-Bowden & Cárdenas 2017; Goldenfeld, Biancalani, & Jafarpour, 2017, Vetsigian, Woese, & Goldenfeld 2006; Woese, 2002). SOR bridges psychological research on fundamental aspects of our human nature such as creativity and our proclivity to reflect on ideas from different perspectives, with the literature on evolutionary approaches to cultural evolution that aspire to synthesize the behavioral sciences much as has been done for the biological scientists. The current chapter is complementary to this effort, but less abstract; it attempts to ground the theory of cultural evolution in terms of cognitive transitions as suggested by archaeological evidence.

By the term *archaeological evidence* we mean the 'material correlates' or 'precipitates' of behavior. While artifacts were once treated as indicators of 'progress', more contemporary approaches, including the approach taken here, seek to reveal the cognitive conditions responsible for artifacts (and other material precipitates of behavior; Haidle, 2009, Wragg-Sykes, 2015). Note that although our theoretical approach is founded on evolutionary principles, it is not what is often referred to as 'evolutionary psychology' (Cosmides & Tooby, 1992; Sell et al, 2009) which focuses on biological underpinnings of cultural evolution as opposed to the impact of culture as an evolutionary process unto itself. Our approach is more aligned to other evolutionary approaches to the general question of how modern cognition arose, such as those of Wynn and Coolidge (e.g., Wynn et al, 2017) highlighting developmental psychology and Bruner (e.g. 2010), highlighting palaeoneurology.

Our use of the term 'transition' in the title is intensional (for as Straus, 2009, observes the term 'transition' is sometimes used too casually). We begin the chapter with a discussion of the concept of evolutionary transition, for the 'unpacking' of this term could be of explanatory value with respect to archaeological change and its cognitive underpinnings.

## Evolutionary Transitions

Transitions are common in evolution (Szathmary & Maynard Smith, 1995). Nonlinear interactions between different information levels (e.g., genotype, phenotype, environment, and even developmental characteristics) often give rise to emergent outcomes that generate discontinuities (Galis & Metz, 2007). Research into the mechanisms underlying evolutionary transitions have made headway into explaining the origins of new varieties of information organization and unpacking terms such 'adaptation due to natural selection', aiming "to analyze trends of increasing complexity" (Griesmer, 2000). Szathmary and Maynard-Smith's account of the eight major transitions in the history of life remains widely accepted today (Szathmary 2015, Calcott & Sterelny, 2011), with other transitions continue to be identified, including the evolution of new sexes (Parker, 2004), and new varieties of ant agriculture (Schultz & Brady, 2008), animal individuality (Godfrey-Smith 2011), metabolism and cell structure (DeLong et al., 2010), technology (Geels, 2002) and hominin socialization (Foley & Gamble 2009).

Research on the dynamics (e.g., rates and types) of evolutionary transitions shows that despite their diversity they exhibit common features: they are (1) rare, (2) involve new levels of organization of information, (3) followed by diversification, and (4) incomplete (Wilson 2010). Szathmary and Maynard-Smith include the transition from "primate societies to human societies " as part of their "Extended Evolutionary Synthesis" (Wilson, 2010), but this synthesis was formulated just prior to the rise of explicitly evolutionary approaches to modern cognition.

We suggest that the theory of evolutionary transitions can provide a useful framework for understanding the cognitive changes culminating in behavioral modernity (BM). In cognitive evolution, evidence of significant change might be stretched out over time and space for many reasons, such as lag between initial appearance and demic diffusion, ambiguities in the archaeological and fossil records. In this chapter we explore two such transitions. The first is the origin of a richer, post-*Pan*, post-*Australopithecine* culture as early as 2.2 million years ago (Harmand et al., 2015). The second is the explosion of creative culture in the Middle/Upper Paleolithic.

## Origin of Post-*Pan*, Post-*Australopithecine* Culture: A First Cognitive Transition Archaeological and Anthropological Evidence

The minds of *Australopithecus* and earliest *Homo* appear to have been anchored to the present moment of concrete sensory perceptions, i.e., the "here and now". Simple stone (and some bone and antler) implements indicates that they encoded perceptions of events in memory, thereby supplying "timely information to the organism's decision-making systems" (Klein et al., 2002, page 306)—but had little voluntary access to memories without external cues. The upshot was minimal innovation and artifact variation.

This is evident in the early archaeological record, beginning with stone tools from Lomekwi 3 West Turkana, Kenya, 3.3 mya (Harmand et al., 2015), and characterized by opportunism in highly restricted environments (Braun et al., 2008). Tools were technologically on par with those of modern chimpanzees (Byrne & Russon, 1998; Blackwell & d'Errico, 2001; see Read [2008] and Fuentes [2015] for cognitive considerations of chimpanzee toolmaking). These tools lack evidence of symbolism (d'Errico et al., 2003), and were transported relatively short distances across landscapes (Potts, 2012). While nut-cracking and other simple tool use outside Homo may involve sequential chaining of actions, (and thus sequential chaining of the mental representations underlying these actions), outside Homo, sequential processing did not occur reliably enough to cross the threshold for abstract thought (see Gabora & Steel, 2017 for a mathematical model of this). Thus, it appears that early Homo could not invent or refine complex actions, gestures, or vocalizations, and their ability to voluntarily shape, modify, or practice skills and actions was at best minimal.

Early *Homo* evolved into several forms, including *H. erectus*, dating between 2.8 - 0.3 million years ago (Villmoare et al., 2015), and there was a shift away from biology and towards culture as the primary means of adaption in this lineage, attended by significant cultural elaboration. Having expanded out from Africa as early as 2 mya, *Homo* constructed tools involving more production steps and more varied raw materials (Haidle, 2009), imposed symmetry on tool stone (Lepre et al., 2011), used and controlled fire (Goren-Inbar et al., 2004), crossed stretches of open water up to 20 km (Gibbons 1998), ranged as far north as latitude 52∘ (Preece et al., 2010), revisited campsites possibly for seasons at a time, built shelters (Mania & Mania 2005), transported tool stone over greater distances than their predecessors (Moutsou, 2014), and ranked moderately high among predators (Plummer, 2004). While the cranial capacity of *Homo erectus* was approximately 1,000 cc—about 25% larger than that of *Homo habilis*, at least twice as large as that of living great apes, and 75% that of modern humans (Aiello, 1996)—brain volume alone cannot explain these developments. It is widely thought that these signs of a culture richer than that of *Pan* or *Australopithecus* c. 1.7 mya reflect a significant transition in cognitive and/or social characteristics.

## Early Ideas about What Caused this First Transition

We can take as a starting point Donald's (1991) theory of cognitive evolution as it was a breakthrough that paved the way for much of what followed. Because the cognition of *Homo habilis* was primarily restricted to whatever episode one was currently experience, Donald refers to it as an *episodic mode* of cognition. He proposed that the enlarged cranial capacity brought about the onset of what he calls a *self-triggered recall and rehearsal loop*, which we abbreviate STR. STR enabled hominins to voluntarily retrieve stored memories independent of environmental cues (sometimes referred to as 'autocuing') and engage in *representational redescription* and the refinement of thoughts and ideas. This was a fundamentally new mode of cognitive functioning which Donald referred to as the *mimetic mind* because it could 'mime' or act out past or possible future events, thereby not only temporarily escaping the present, and communicating that escape to others.

STR also enabled attention to be directed away from the outside world toward ones' internal representations, paving the way for abstract thought. We use the term *abstract thought* to refer to the reprocessing of previous experiences, as in counterfactual thinking, planning, or creativity, as opposed to direct perception of the 'here and now' (for a review of abstract thought, see Barsalou, 2005). Note that in much of the cultural evolution literature, abstract thought and creativity, if mentioned at all, are equated with individual learning, which is thought to mean 'learning for oneself' (e.g., Henrich & Boyd, 2002; Mesoudi, Whiten & Laland, 2006; Rogers, 1988). However, they are not the same thing; *individual learning* deals with obtaining pre-existing information from the environment through non-social means (e.g., learning to differentiate different kinds of trees). The information does not change form just because the individual now knows it. In contrast, *abstract thought* involves the *reprocessing* of internally sourced mental contents, such that they are *in flux*, and when this results in the generation of useful or pleasing ideas, behavior, or artifacts that did not previously exist, it is said to be *creative*. Indeed, there is increasing recognition of the extent to which creative outcomes are contingent upon internally driven incremental/iterative honing or reprocessing (Basadur, 1995; Chan & Schunn, 2015; Feinstein, 2006; Gabora, 2017).

Note that Donald's explanation focuses on neither technical nor social abilities but on a cognitive ability that facilitated both. STR enabled systematic evaluation and improvement of thoughts and motor skills such that they could be adapted to new situations, resulting in voluntary rehearsal and refinement of actions and artifacts. STR also broadened the scope of social activities to include re-enactive play and pantomime.

## Proposed Theory of Cognitive Underpinnings of this First Transition

Leaving aside alternatives to Donald's proposal until the end of this section, we note for now that although Donald's theory seems reasonable, it does not explain why larger brain size enabled STR. In what follows, we contextualize Donald's (1991) schema in research on the nature of associative memory. We will ground the concept of STR in a neural level account of the mechanisms underlying cognitive flexibility and creativity (Gabora, 2010; Gabora &

Ranjan, 2013).

We start by summarizing a few well-known features of associative memory. Each neuron is sensitive to a primitive stimulus attribute, or microfeature, such as lines of a particular orientation, or sounds of a particular pitch. Items in memory are *distributed* across cell assemblies of such neurons; thus each neuron participates in the encoding of many items. Memory is also *content-addressable:* there is a systematic relationship between the content of an item and the neurons that encode it; therefore, items that share microfeatures may be encoded in overlapping distributions of neurons.

While in and of itself increased brain volume cannot explain the origin of BM, we suggest that larger brains enabled a transition from coarse-grained to more fine-grained memory. The smaller the number of neurons a brain has to work with, the fewer attributes of any given memory item it can encode, and the less able it is to forge associations on the basis of shared attributes. Conversely, the evolution of finer-grained memory meant that memory items could be encoded in more detail, i.e., distributed across larger sets of cell assemblies containing more neurons. Since memory organization was content addressable, this meant more ways in which distributed representations could meaningfully overlap, and greater overlap enabled more routes by which one memory could evoke another.

This in turn made possible the onset of STR, and paved the way for the capacity to engage in recursive recall and streams of abstract thought, and a limited form of insight (Gabora, 2002, 2010; Gabora & Ranjan, 2013). As a simple example, the reason that the experience of seeing a leaf floating on a lake could potentially play a role in the invention of a raft is that both experiences involve overlap in the set of relevant attribute "float on water", and thus overlap of activated cell assemblies.

Items in memorys could now be reprocessed until they achieved a form that was acceptably consistent with existing understandings or sufficiently enabled goals and desires to be achieved (Gabora, 1999. This scenario provides a plausible neural-level account of Donald's (1991) proposal that abstract thought was a natural consequence of possessing a self-triggered recall and rehearsal loop, made possible by an increase in brain size.

**Comparison with Other Theories**
Mithen's (1996) model features the accumulation and overlap of different intelligence modules. Although in its details his model runs rather counter to much current thinking including our own, his thoughts on cognitive fluidity and creativity influenced the model proposed here.

Some theories attribute this transition to social factors. Foley and Gamble (2009) emphasize enhanced family bonding and the capacity for a more focused style of concentration, further enhanced by controlled use of fire by at least 400,000 years ago. Wiessner (2014) suggested that fire not only enabled the preparation of healthier food, but by providing light after dark, facilitated playful and imaginative social bonding. Others emphasize an extrication from biologically based to culturally based kinship networks (Leaf & Read, 2012; Read, Lane & van der Leeuw, 2015). We believe that these social explanations are correct but that they have their origin in cognitive changes, which altered not only social interactions but interactions with other aspects of human experience as well.

Our proposal bears some resemblance to Hauser et al.'s (2002) suggestion that the capacity for recursion is what distinguishes human cognition from that of other species, Penn et al's (2008) concept of relational reinterpretation, and Read's (2009) claim that recursive reasoning with relational concepts made possible a conceptually based system of social relations but may have evolved alongside non-social activities such as toolmaking. However, our proposal goes further, because it grounds the onset of recursive reasoning in a specific cognitive transition. While Read suggests that recursive reasoning was made possible by larger working memory, we argue that larger working memory in and of itself is not useful; it must goes hand-in-hand with—and indeed is a natural byproduct of—a finer-grained memory. As an illustrative example, let us suppose that a hominid with a coarse-grained memory increased its working memory from being able to think only of one thing at a time (e.g., a leaf) to two (e.g., a leaf and the moon). This would generally be a source of confusion. However, if it held only one thing in mind at a time but encoded it in richer detail (e.g., incorporating attributes of a leaf such as 'thin', 'flat', and so forth), it could forge meaningful associations with other items based on these attributes (e.g., other thin, flat things).

Our proposal also bears some similarity to Chomsky's (2008) suggestion that this transition reflects the onset of the capacity for a 'merge' operation. 'Merge' is described as the forging of associations based on their global similarity, i.e., between items that are highly similar or that co-occur in space or time. In contrast, for STR the memory must be sufficiently fine-grained (i.e., items must be encoded in sufficient detail) that the associative process can operate on the basis of *specific attributes* to which specific neurons are tuned. STR can forge associations between items that are related by as few as a single attribute, and do so recursively such that the output of one such operation is the input for the next, and reliably, such that encodings are modified in light of each other in the course of streams of thought (Gabora, 2002, 2013, 2017; detailed examples, such as the invention of a fence made of skis on the basis of the attributes 'tall', 'skinny' and 'sturdy' [Gabora, 2010], and the generation of the idea of a beanbag chair on the basis of the single attribute 'conforms to shape' [Gabora, 2018], are provided elsewhere). Thus our proposal (but not 'merge') offers a causal link between brain size and cognitive ability, i.e., more neurons means they can be tuned to a broader range of attributes and thereby form more associations on the basis of shared attributes.

**Creative Culture in Middle/Upper Paleolithic: A Second Cognitive Transition**
**Archaeological and Anthropological Evidence**
The African archaeological record indicates another significant cultural transition approximately 100,000 years ago that shows many material correlates of BM. Though there is no single definitive indicator of BM (d'Errico et al., 2005; Shea 2011), it is generally thought to involve (a) a radical proliferation of tool types that better fit tools to specific tasks (McBrearty & Brooks, 2000), (b) elaborate burial sites indicating ritual (Hovers, Ilani, Bar-Yosef, & Vandermeersch, 2003) and possibly religion (Rappaport, 1999), (c) artifacts indicating personal symbolic ornamentation (d'Errico et al., 2009), (d) 'cave art', i.e., representational imagery featuring depictions of animals (Pike et al., 2012) and human beings (Nelson, 2008), (e) complex hearths and highly structured use of living spaces (Otte, 2012), (f) calorie-gathering intensification that included widespread use of aquatic resources (Erlandson, 2001), and (g) extensive use of bone and antler tools, sometimes with engraved designs. BM extended across Africa after 100,000 years ago, and was present in Sub-Himalayan Asia and Australasia over 50,000 years ago (Mulvaney & Kamminga, 1999) and Continental Europe thereafter (Mellars, 2006).

It is uncertain whether this archaeological record reflects a genuine transition resulting in BM because claims to this effect are based on the European Paleolithic record, and largely exclude the lesser-known African record (Fisher & Ridley, 2013). Artifacts associated with a rapid transition to BM 40,000-50,000 years ago in Europe are found in the African Middle Stone Age tens of thousands of years earlier, pushing the cultural transition more closely into alignment with the transition to anatomical modernity between 200,000 and 100,000 BP. Nevertheless, it seems clear that BM appeared in Africa between 100,000 to 50,000 years ago, and spread to Europe, resulting in displacement of the Neanderthals (Fisher & Ridley, 2013). Despite an overall lack of increase in cranial capacity, the prefrontal cortex, and more particularly the orbitofrontal region, increased significantly in size (Dunbar 1993), in what was likely a time of major neural reorganization (Morgan 2013). *Homo sapiens* could now effectively archive information, and adapt it to different needs and circumstances, making their cultures radically more creative, open-ended, and accumulative than any prior hominin (Mithen, 1998).

This transition is also commonly associated with the origins of complex language. Although the ambiguity of the archaeological evidence makes it difficult to know exactly when language began (Davidson & Noble 1989; Christiansen & Kirby, 2003; Hauser, Chomsky, & Fitch, 2002), it is widely believed—based on stone tool symmetry and complexity of manufacture—that as long ago as c. 1.7 million years Homo used gestural and prelinguistic vocalization communications that would have shared some organizational similarities with modern humans insofar as they differed significantly from other primate communications. The evolution of grammatically- and syntactically-modern language is generally placed (depending on whether one is observing it in Africa, sub-Himalayan Asia or Western Eurasia) after about 100,000 years ago, around the start of the Upper Palaeolithic (Bickerton, 2014; Dunbar, 1993; Tomasello, 1999).

## Proposed Cognitive Mechanism Underlying Second Transition

We propose that the root cause of the cultural explosion of the Middle/Upper Paleolithic was a fine-tuning of the biochemical mechanisms underlying the capacity to spontaneously shift between different modes of thought in response to the situation, and that this is accomplished by varying the specificity of the activated region of memory. The ability to shift between different modes is referred to as *contextual focus* (CF) because it requires the capacity to focus or defocus attention in response to contextual factors (Gabora, 2003), such as a specific a goal, a particular audience, or aspect of the situation, and do so continuously throughout a task. Focused attention is conducive to *analytical thought* (Agnoli, Franchin, Rubaltelli, & Corazza, 2015; Vartanian, 2009; Zabelina, 2018), wherein activation of memory is constrained enough to hone in and mentally operate on only the relevant aspects of mental contents. In contrast, by diffusely activating a wide region of memory, defocused attention is conducive to *associative thought*; it enables more obscure (though potentially relevant) aspects of the situation to be considered. This greatly enhances the potential for insight, i.e., the forging of obscure but useful or relevant connections.

Note that associative thought is useful for breaking out of a rut, but would be risky without the ability to reign it back in; basic survival related tasks may be impeded if everything is reminding you of everything else. Therefore, it would take considerable time to fine-tune the mechanisms underlying the capacity to spontaneously shift between these two processing modes such that one retained the benefits of escaping local minima without running the risk of being perpetually side-tracked. The time needed to fine-tune this could potentially be the explanation for the lag between anatomical and BM.

Once the products of one mode of thought could become 'ingredients' for the other, hominids could reflect on thoughts and ideas not just from different perspectives but at different levels of granularity, from basic level concepts (e.g., rabbit) up to abstract concepts (e.g., animal) and down to more detailed levels (e.g., legs), as well as conceive of their interrelationships. This kind of personal reflection was necessary for, and indeed a precursor to, the need to come up with names for these things, i.e., the development of complex languages. Thus, it is proposed that CF paved the way for not just language but the seemingly diverse collection of cognitive abilities considered diagnostic of BM.

To see how the onset of CF could result in open-ended cultural complexity, recall that associative memory has the following properties: distributed representation, coarse coding, and content addressability. Each thought may activate more or fewer cell assemblies depending on the nature of the task. Focused attention is conducive to analytic thought because memory activation is constrained enough to zero in and operate on key defining properties. Defocused attention, by diffusely activating a broader region of memory, is conducive to associative thought; obscure (but potentially relevant) aspects of the situation come into play (Gabora, 2000, 2010). Thus, thinking of, say, the concept SUN in an analytic mode might bring to mind only the literal sun, in an associative mode of thought it might also bring to mind other sources of heat, or other stars, or even someone with a sunny disposition. Once our hominid ancestors could shift between these modes of thought, tasks requiring either one mode, or the other, or shifting between them in a precisely orchestrated sequence the fruits of one mode become ingredients for the other. This resulted in the forging of a richly integrated creative internal network of understandings about the world and one's place in it, or *worldview,* which we claim made BM possible. Thus, the notion that diffuse memory activation is conducive to associative thought while activation of a narrow receptive field is conducive to analytic thought is not only consistent with the architecture of associative memory, but suggests an underlying mechanism by which CF enabled the ability to both (1) stay task-focused, and (2) deviate from the task to make new connections, as needed. In this view, language did not evolve solely to help people communicate and collaborate (thereby accelerating the pace of cultural innovation); it also helped people think ideas through for themselves and manipulate them in a deliberate, controlled manner. Language facilitated the weaving of experiences into stories, parables, and broader conceptual frameworks, thereby integrating knowledge and experience (see also, Gabora & Aerts, 2009). Instead of staying squarely in the narrow regime between order and chaos, thought could now shift between orderly and chaotic, as appropriate.

Thus, we propose that the emergence of a self-organizing worldview required two transitions. The onset of STR over 2 mya allowed rehearsal and refinement of skills and made possible minor modifications of representations. The onset of CF approximately 100,000 years ago made it possible to forge larger bridges through conceptual space that paved the way for innovations specifically tailored to selective pressures. It enabled a cultural version of what Gould and Vrba (1982) call *exaptation,* wherein an existing trait is co-opted for a new function (Gabora, Scott, & Kauffman, 2013). Exaptation of representations and ideas vastly enhanced the ability to expand the technological and social spheres of life, as well as to develop individualized perspectives conducive to fulfilling complementary social roles. This increase in cognitive variation provided the raw material for enhanced cultural adaptation to selective pressures.

**Comparison with Other Theories**

Our proposal is superficially similar to (and predates – see Gabora 2003) the suggestion that what distinguishes human cognition from that of other species is our capacity for *dual processing* (Evans, 2008; Nosek, 2007). This is the hypothesis that humans engage in (1) a primitive implicit Type 1 mode involving free association and fast 'gut responses', but (2) an explicit Type 2 mode involving deliberate analysis. However, while dual processing makes the split between older, more automatic processes and newer, more deliberate processes, CF makes the split between an older associative mode based on relationships of correlation and a newer analytic mode based on relationships of causation. We propose that although earlier hominids relied on the older association-based system, because their memories were coarser-grained, there were fewer routes for meaningful associations, so there was less associative processing of previous experiences. Therefore, items encoded in memory tended to remain in the same form as when they were originally assimilated; rather than engaging in associative or analytic processing of previously assimilated material, there was greater tendency to leave mental contents in their original form and instead focus on the present. Thus, while dual processing theory attributes abstract, hypothetical thinking to the more recent Type 2 mode, according to our theory abstract thought is possible in either mode but differs in character in the two modes: logically constructed arguments in the analytic mode versus flights of fancy in the associative mode. Our theory is rooted in a distinction in the creativity literature between (1) associative divergent processes said to predominate during idea generation, and (2) analytic convergent processes said to predominate during the refinement and testing of an idea (Finke, Ward, & Smith, 1992; see also Sowden, Pringle, & Gabora [2014], for a comparison of theories of creativity; and Gabora, 2018 for the distinction between associative versus divergent thought).

Mithen (1996) proposed that BM came about through the integration of previously-compartmentalized intelligence modules that were specialized for natural history, technology, socialization and language. This integration, he says, enabled *cognitive fluidity*: the capacity to combine concepts and adapt ideas to new contexts, and thereby explore, map, and transform conceptual spaces across different knowledge systems. A related proposal emphasizes the benefit of cognitive fluidity for the capacity to make and understand analogies (Fauconnier & Turner, 2002). Our explanation is consistent with these but goes beyond them by showing how conceptual fluidity would arise naturally as a function of the capacity to, when appropriate, shift to a more associative processing mode.

There are many versions of the theory that BM reflects the onset of complex language complete with recursive embedding of syntactic structure (Bickerton & Szathmáry, 2009; Carstairs-McCarthy, 1999), which enabled symbolic representation and abstract thought (Bickerton, 2014; Deacon, 1997), narrative myth (Donald, 1991), enhanced communication, cooperation, and group identity (Voorhees, Read & Gabora, in press). Our proposal is consistent with the view that complex language lay at the heart of BM, but because STR followed by CF would have enabled hominids to not just recursively refine and modify thoughts but consider them from different perspectives at different hierarchical levels, it set the stage for complex language.

Since, as explained above, we see evidence of recursive reasoning well before BM, our framework is inconsistent with the hypothesis that onset of recursive thought enabled mental time travel, distinctly-human cognition, and BM (Corballis, 2011; see also Suddendorf et al., 2009). Nevertheless, we suggest that the ability to shift between different modes of thought using CF would have brought on the capacity to make vastly better use of it. The proposal that BM arose due to onset of the capacity to model the contents of other minds, sometimes referred to as 'Theory of Mind' (Tomasello, 2014) is somewhat underwritten by recursion, since the mechanism that allows for recursion is required for modeling the contents of other minds (though in this case the emphasis is on the social impact of recursion, rather than the capacity for recursion itself). Our proposal is also consistent with explanations for BM that emphasize social-ecological factors (Foley & Gamble, 2009; Whiten, 2011), but places these explanations in a broader framework by suggesting a mechanism that aided not just social skills but other skills (e.g., technological) as well. While most of these explanations are correct insofar as they go, we suggest that they do not get to the root of the matter. As Carl Woese wrote of science at large "...sometimes [there is] no single best representation... only deeper understanding, more revealing and enveloping representations," (2004, page 173). We propose that the second cognitive transition necessary for cumulative, adaptive, open-ended cultural evolution was the onset of CF, because once hominids could adapt their mode of thought to their situation by reflecting on mental contents through the lenses of different perspectives, at different levels of analysis, their initially fragmented mental models of their world could be integrated into more coherent representations of their world—i.e., worldviews. This facilitated not just survival skills, conceptual fluidity, and creative problem-solving, but also social exchange and the emergence of complex social structures.

In short, the explanation proposed here is the only one we know of that grew out of a synthesis of archaeological and anthropological data with theories and research from both psychology and neuroscience, and it appears to underwrite explanations having specifically to do with dual processing, conceptual fluidity, language, social interaction, or theory of mind.

## A Tentative Genetic Basis for Contextual Focus

The hypothesis that BM arose due to the onset of contextual focus (CF, or the capacity to shift between different modes of thought), leads to the question: what caused CF? In this section we explore a possible genetic basis for CF. First, we provide historical context by reviewing both studies that implicated genes such as FOXP2 in the origins of language, and the evidence that caused this explanation to fall out of favor. Next, we review anthropological and archaeological evidence that the coming into prominence of creative and cognitive abilities (including but not limited to those that involve language) coincides with the evolutionary origins of FOXP2. Then we synthesize these literatures in a new explanation of the role of FOXP2 in language and cognition.

FOXP2, a transcription gene on chromosome 7 (Reimers-Kipping, Hevers, Paabo, & Enard 2011), regulates the activity of other genes involved in the development and function of the brain (Fisher & Ridley 2013; Kovas & Plomin 2006). Among other functions, FOXP2 plays a role in the functioning of the motor cortex, the striatum, and the cerebellum, which controls fine motor skills (Liegeois, Morgan, Connelly, & Vargha-Khadem 2011). There are findings of familial resemblance for specific components of linguistic competence and suggestions of a genetic basis for such competences (e.g., Kovac, Gopnik, & Palmour 2002).

The finding that a mutation in FOXP2 was associated with language comprehension and production in a family known as the KE Family (Lai, Fisher, Hurst, Vargha-Khadem, & Monaco 2001) led to the proposal that the gene plays a key role in language acquisition. The family was diagnosed with Specific Language Impairment (SLI), a severe deficit in language development that exists despite adequate educational opportunity and normal nonverbal intelligence (Lai et al. 2001; Morgan 2013). Those afflicted with SLI exhibit deficits in speech comprehension as well as verbal dyspraxia – a severe difficulty controlling the movement and sequencing of orofacial muscles required for the articulation of fluent speech (Lalmansingh, Karmakar, Jin, & Nagaich, 2012). Thus FOXP2 became prematurely known as the "language gene" (Bickerton, 2014).

Evidence that FOXP2 is actively transcribed in brain areas where mirror neurons are present suggested that FOXP2 makes language possible at least partially through its effect on the capacity to imitate (Corballis, 2004). This hypothesis was strengthened by findings that mirror neurons play a key role in language development (Arbib, 2011). However, in order to imitate the language of others there must already be others using language. Therefore, it is difficult to fully account for the origin of language by positing that FOXP2 affects language by way of its effect on imitation. Moreover, this cannot explain the existence of defects associated with FOXP2 that involve neither language nor imitation. Finally, other species imitate but do not exhibit grammatical, syntactically rich language.

The recognition that FOXP2 was not the "language gene" generated sober discussion about the simplicity of single gene explanations for complex traits. Indeed, as a transcription gene, FOXP2 only has an indirect effect on neural structure or behavior (Reimers-Kipping et al. 2011). Evidence that FOXP2 affects abilities other than language, such as cognitive fluidity, and even IQ (Kurt, Fisher, & Ehret 2012; Lai et al. 2001), and plays a broad role in the modulation of neural plasticity (Fisher & Scharff 2009), suggest that the neurological basis of FOXP2 deficits lie in the structural and functional abnormalities of cortico-striatal and cortico-cerebellar circuitries, which are important for learning, memory, and motor control, not language exclusively. The structure of language is now widely believed to have come about through ontogenetic human learning and processing mechanisms (Christiansen & Chater 2008), with diversity of linguistic organization the rule rather than the exception, and instances of universality reflecting stable engineering solutions satisfying multiple design constraints rather than natural selection (Evans & Levinson 2009).

Language capacity is now widely thought to be polygenic, i.e., affected by multiple genes (Chabris et al., 2012; Kovas & Plomin, 2006), transcription genes such as FOXP2 are often pleiotropic, i.e., affecting multiple traits. It has been firmly established that a small perturbation (such as a mutation) can percolate through a system resulting in widespread, large-scale change, a phenomenon known as self-organized criticality (Bak, Tang, & Wiesenfeld, 1987), and self-organized criticality is particularly widespread in regulatory genes (Kauffman, 1993). Thus, the possibility that FOXP2 plays a role in cognition that extends beyond language is worthy of consideration.

While it is clear that FOXP2 is important to language capacity in modern human populations, a recent re-examination of the gene in a large, global, sample concludes that there is "no evidence that the original two amino acid substitutions were targeted by a recent sweep limited to modern humans <200 kya...[and furthermore] recent natural selection in the ancestral Homo sapiens population cannot be attributed to the FOXP2 locus and thus Homo sapiens' development of spoken language" (Atkinson et al. 2018, page 9). However, as noted above, FOXP2 is only one of one of several genes strongly implicated in conditioning language capacity; Mozzi et al. (2016) indicate that nine other genes are found, in various states, to compromise skills listed as language, reading, and speech, thus effecting both cognition (language and reading) and motor control (speech). Furthermore, the evolutionary history of these genes is under investigation, and promises to enrich our understanding of the evolution of the cognition responsible for language and, by extension, behavioral modernity.

Given the evidence that FOXP2 plays a role in the evolution of complex cognitive abilities including language, but this relationship is not solely mediated through its effects on the capacity for imitation, it seems reasonable to propose that FOXP2 and/or other, associated genes or transcription factors enabled CF. Interestingly, the appearance of anatomically modern humans in the fossil record as early as 200,000 years ago coincides with accelerated FOXP2 evolution (Corballis 2004b; Lai et al. 2001). It has been proposed that the appearance of anatomically modern humans was due to amino acid substitutions that differentiate the human FOXP2 gene from that of chimpanzees (Enard et al. 2002; Zhang et al. 2002). Despite new challenges to the evolutionary history of FOXP2 (see discussion above) the preponderance of evidence currently suggest that within the last 200,000 years FOXP2 underwent at least two human-specific mutations, at least one of which occurred within the last 100,000 years (Lai et al. 2001; Morgan 2013). This chronologically aligns modifications of FOXP2 with the onset of not just anatomical modernity but also BM. Hence we propose that the Paleolithic transition to BM reflects a mutation to FOXP2 and/or its molecular associates that facilitated fine-tuning of the capacity to spontaneously shift between associative and analytic modes (CF) depending on the situation by varying the specificity of the activated memory region.

We propose that the kickoff point for the explosion of creativity and onset of language in the Middle/Upper Paleolithic was human-specific amino acid substitutions to FOXP2 and/or its associates changed modifications of basal ganglia neurons that contribute to cognitive flexibility. These neurons have longer dendrite length and greater synaptic plasticity in humans compared to chimpanzees. These changes enhanced the efficiency of neural cortico-basal ganglia circuits, enabling individuals to spontaneously adjust to what extent the details of a given item in memory contributed to the flow of thought. For example, if not just the salient or defining aspects of a particular representation are activated but also peripheral aspects, then these peripheral features could trigger remote associations, thereby increasing conceptual fluidity. Our theory is consistent with the proposal that mutations in FOXP2 and/or its associates underwrote the capacity for both the gestures and grammars associated with BM (see Vicario 2013).

In short, we suggest that while FOXP2 is not the language gene it may have a broad and identifiable influence on cognition by enabling the capacity for CF. This is not incompatible with Crow's (2012) proposal that the Protocadherin11XY gene pair played a key role in establishing cerebral asymmetry and enabling complex language. However, our proposal is compatible with evidence that FOXP2 and related genes paved the way for not just language but other features of BM as well. By tuning the mode of thought to match the needs of the present moment, CF allowed information to be processed at different degrees of granularity, and from different perspectives. Thus it is by way of CF that hominins became able to combine actions and words into an infinite variety of cultural outputs, and respond to changing selective pressures. Individuals could engage in convergent thought for well-defined tasks, but shift to divergent thought when they were stuck, or when they wanted to express themselves or explore aesthetic possibilities. This enabled them to connect ideas in new ways, resulting in advanced tools, elaborate burials, and different forms of creative expression, including art and jewelry.

## Discussion

This chapter outlined a theory of how the the uniquely human capacity for collectively generated, open-ended, adaptive cultural evolution could have come about. Although change occurred in a mosaic fashion in the Homo lineage over more than two million years, two significant evolutionary transitions stand out. Thus, we propose that the distinctive rich symbolism and grammatically complex language of the genus *Homo* reflect two evolutionary transitions brought about by novel forms of cognitive information processing.

First, the larger brain of *H. erectus* resulted in finer grained memory with detailed representations, paving the way for rehearsal of actions, refinement of skills, novel associations between closely related items in memory. This enabled STR, escape from episodic proximity, representational redescription, minor improvements in cultural outputs, and a "cultural ratcheting" that expanded the capacity for open-ended cultural evolution.

The second transition occurred approximately 100,000 BP, a period associated with the origins of art, science and religion (Mithen, 1998). We suggest that newly-evolved basal ganglia circuits enabled onset of contextual focus: the ability to shift between convergent and divergent modes of thought, enabling hominins to process information from different perspectives and at multiple levels of detail. Hominins could now put their own spin on the ideas of others, adapting them to individual needs and tastes, leading to cumulative innovation. Thoughts, impressions, and attitudes could be modified by thinking about them in the context of each other, and they could be woven into an integrated "worldview" that defines who we are in relation to the world. This allowed the capacity for self-expression, creating an environment conducive to the emergence of complex language, including grammar, recursion, word inflections, and syntactical structure, as well as comprehension.

This theory is consistent with findings that FOXP2 is associated with cognitive abilities that do not involve language, and with findings that non-language creative abilities arose at approximately the same time as complex language (Chrusch & Gabora, 2014). It is also consistent with findings that despite the existence of sophisticated cognitive abilities in other species such as birds (Emery, 2016), we alone exhibit cumulative cultural evolution. Cumulative cultural evolution may involve the 'recycling' of cortical maps such that cultural innovations invade evolutionarily older brain circuits and inherit some of their structural constraints (Dehaene, 2005; Lieberman, 2016).

Elsewhere we provide support for the proposed two-transition scenario obtained using an agent-based model of cultural evolution (Gabora & Smith 2018). We note that the origins of BM are currently being rethought in light of wide dissatisfaction with an archaic 'trait-list' approach to its understanding (Ames, Riel-Salvatore, & Collins, 2013) and with nonlinear models of multifaceted cultural evolutionary change (Mesoudi, 2009; McDowell, 2013). The transitions to possession of the cognitive capacities that we propose made BM possible—STR and CF—exhibit Wilson's (2010) defining characteristics of evolutionary transitions, i.e., they are rare, incomplete (did not 'throw a switch' resulting in immediate 'turning on' of BM), and involved new levels of organization. The increased sociality implied by the onset of STR and CF also meets Wilson's expectation that evolutionary transitions drive "the suppression of fitness differences within groups, causing between-group selection to become the primary evolutionary force" (Wilson, 2010:135). It is our hope that the proposed theory of cognitive evolution reflects an emerging 'Extended Evolutionary Synthesis' (Smith & Ruppell, 2011, Smith, Gabora, & Gardner-O'Kearny *in press*; Woese, 2004). We suggest that the origins of BM be considered an evolutionary transition that culminated in new varieties of information both within the mind and in artificial memory systems external to it, giving way to new social arrangements and paving the way for the complex cultural systems in which we are presently immersed.

## Acknowledgments

This work was supported by a grant (62R06523) to the first author from the Natural Sciences and Engineering Research Council of Canada.